\documentclass[a4paper]{jpconf}
\usepackage{graphicx}
\usepackage{hyperref}
\usepackage{epstopdf}
\usepackage{amsmath}

\newcommand{\ai}{a_{i}}
\newcommand{\Ui}{\Upsilon_{i}}
\newcommand{\gi}{\gamma_{i}}
\newcommand{\Hi}{H_{i}}
\newcommand{\Ti}{T_{i}}
\newcommand{\si}{s_{i}}

\newcommand{\Tkd}{T_{\rm kd}}

\newcommand{\Tstd}{T_{\rm kd, std}}

\begin{document}

\title{Analytic expressions for the kinetic decoupling of WIMPs}

\author{Luca Visinelli\footnote[1]{Talk based on L.~Visinelli and P.~Gondolo, \href{http://journals.aps.org/prd/abstract/10.1103/PhysRevD.91.083526}{Phys.\ Rev.\ D {\bf 91} (2015) 8, 083526} [\href{http://arxiv.org/abs/1501.02233}{astro-ph/1501.02233}].}}

\address{Department of Physics and Astronomy, University of Bologna, Via Zamboni, 33 - 40126 Bologna}

\ead{luca.visinelli@studio.unibo.it}

\begin{abstract}
We present a general expression for the values of the average kinetic energy and of the temperature of kinetic decoupling of a WIMP, valid for any cosmological model. We show an example of the usage of our solution when the Hubble rate has a power-law dependence on temperature.
\end{abstract}

\section{Introduction}
Despite the various astrophysical observations in support of its existence~\cite{komatsu, planck}, the nature of dark matter still remains an open question. Of the various candidates for dark matter, one of the most compelling is the Weakly Interacting Massive Particle (WIMP)~\cite{zwicky, kolb_book, jungman1996, bertone2005, kuhlen2012}, with a mass ranging from a few GeV to $10{\rm~TeV}$. In fact, when the WIMP annihilation rate falls below the Hubble expansion rate, the chemical equilibrium between WIMPs and the primordial plasma is no longer maintained, and the number of WIMPs per comoving volume naturally fixes to the value required for explaining the present abundance of cold dark matter. Although chemical equilibrium at this stage is no longer maintained, kinetic equilibrium between dark matter and the plasma is still achieved through a high momentum exchange rate~\cite{hofmann, chen, berezinsky, green, bertschinger, bringmann, kasahara, bi, gondolo}. Eventually, when the Hubble rate equates the scattering process rate, WIMPs kinetically decouple from the plasma and flow with a given free-streaming velocity. This velocity sets the lowest value for the size of protohalos, which determines the subsequent evolution of primordial structures~\cite{schmid, boehm, loeb, profumo, gondolo_gelmini, aarssen, cornell}. In particular, Bringmann~\cite{bringmann} defined the temperature of the kinetic decoupling $T_{\rm kd}$ in the standard cosmological scenario, while Gelmini and Gondolo~\cite{gondolo_gelmini} defined $T_{\rm kd}$ in the Low-Temperature Reheating (LTR) cosmology following a dimensionality reasoning.

We present a full solution of the evolution equation governing the process of the kinetic decoupling, and we generalize the definition of the temperature of kinetic decoupling and the average kinetic energy of WIMPs in a generic non-standard cosmological model.

\section{General solution of the temperature equation for Dark Matter in a thermal bath} \label{General equation}

The scattering process between plasma at temperature $T$ and WIMPs of mass $M_\chi \gg T$ is a Brownian motion in momentum space, with momentum transfer related to the number $N_e$ of collisions required to change the momentum by $p$ as $p = \sqrt{N_e}\,\Delta p$. Since $p \sim \sqrt{M_\chi\,T}$ is much larger than the average momentum transfer $\Delta p \sim T$, the number of collisions required to appreciably change the momentum of WIMP is $N_e = (p/\Delta p)^2\sim M_\chi/T \gg 1$. The momentum exchange rate $\Gamma$ is suppressed with respect to the elastic collision rate $\Gamma_{\rm el}$ by a factor $T/M_\chi$. Thermal decoupling of WIMPs occurs at a temperature $\Tkd$ approximatively given by $H(\Tkd) \sim \Gamma$, where $H = H(T)$ is the Hubble expansion rate at temperature $T$. Thermal decoupling of a heavy dark matter particle with $M_\chi \gg T$ and with small momentum transfer per collision $\Delta p \ll p$ is described by a Fokker-Planck equation for the dark matter particle occupation number $f_\chi = f_\chi({\bf p}_\chi)$~\cite{hofmann, berezinsky, green, bertschinger, bringmann, kasahara, gondolo},
\begin{equation} \label{eq:boltzmann}
\frac{\partial f_\chi}{\partial t} - H(T)\,{\bf p}_\chi\,\cdot\,\frac{\partial f_\chi}{\partial {\bf p}_\chi} = \gamma(T)\, \frac{\partial}{\partial {\bf p}_\chi}\,\cdot\,\left({\bf p}_\chi\,f_\chi\,(1\pm f_\chi) + M_\chi\,T\, \frac{\partial f_\chi}{\partial {\bf p}_\chi}\right),
\end{equation}
where $\gamma(T)$ is a monotonically increasing function with $T$.

Defining the WIMP kinetic temperature $T_\chi$ as 2/3 of the average kinetic energy of the dark matter particle,
\begin{equation}
T_\chi = \frac{2}{3}\,\int \frac{{\bf p}_\chi^2}{2M_\chi}\,f_\chi({\bf p}_\chi)\,d^3{\bf p}_\chi,
\end{equation} 
and defining the function
\begin{equation} \label{def_f}
\Upsilon(T) = \frac{\gamma(T)}{H(T)},
\end{equation}
the Fokker-Planck Eq.~(\ref{eq:boltzmann}) in the approximation $1 \pm f_\chi \approx 1$ is rewritten as~\cite{kasahara, gondolo, visinelli}
\begin{equation} \label{mastereq}
a\,\frac{dT_\chi}{da} + 2\left[1+\Upsilon(T)\right]\,T_\chi = 2\Upsilon(T)\,T.
\end{equation}
We solve Eq.~(\ref{mastereq}) in terms of analytic expressions for a generic cosmological model, with the boundary condition that the temperature be $T_i$ for a given scale factor $a_i$, to obtain
\begin{equation} \label{sol2}
T_\chi(a) = T_i\,\left(\frac{a_i}{a}\right)^2\,e^{s(a) - s(\ai)} + \frac{2}{a^{2}}\,\int_{a_i}^a\,e^{s(a) - s(a')}\,\Upsilon(a')\,T(a')\,a'\,da',
\end{equation}
where
\begin{equation}\label{def_s}
s(a) = 2\int^a\,\Upsilon(a')\, \frac{da'}{a'}.
\end{equation}
The solution obtained satisfies the behavior in the ``tight coupling'' limit $\gamma(T) \gg H(T)$ as $a\,T_\chi = {\rm constant}$,  and in the ``decoupled'' limit  $\gamma(T) \ll H(T)$ as  $a^2\,T_\chi = {\rm const}$.

\subsection{Temperature of kinetic decoupling} \label{Temperature of kinetic decoupling}

The temperature of kinetic decoupling $\Tkd$ expresses the temperature of the plasma at which the kinetic decoupling of WIMPs occurs. Here, we use the definition~\cite{gondolo, visinelli},
\begin{equation} \label{relation}
\gamma(\Tkd) = H(\Tkd),
\end{equation}
where $H(\Tkd)$ is the Hubble expansion rate when WIMPs decouple kinetically from the primordial plasma. In the literature, different definitions of the temperature of kinetic decoupling can be found.

\section{Power-law cosmological model}\label{Power-law models}

\subsection{General relations for a cosmological model}

We assume that the Hubble rate depends on temperature as
\begin{equation} \label{H_powerlaw}
H(T) = \Hi\,\left(\frac{T}{\Ti}\right)^\nu,
\end{equation}
where $\nu$ is a positive constant, and $\Ti$ and $\Hi$ are the temperature of the plasma and the expansion rate at the time at which we start considering the cosmological model. We also set
\begin{equation} \label{T_a_powerlaw}
a^{\alpha} \, T   = {\rm const}.
\end{equation}
Equating Eqs.~(\ref{H_powerlaw}) and~(\ref{T_a_powerlaw}), we obtain the relation
\begin{equation} \label{H_a_powerlaw}
H(a) = \Hi\,\left(\frac{\ai}{a}\right)^{\nu\,\alpha},
\end{equation}
where $\ai$ is the scale factor at temperature $\Ti$. 
Notice that, in the radiation-dominated cosmology for which $\nu=2$ and $\alpha=1$, the temperature of the plasma drops as $T \propto a^{-1}$, while the WIMP temperature drops at a faster rate $T_\chi \propto a^{-2}$. For the momentum relaxation rate $\gamma(T)$ we assume a power-law function of the form
\begin{equation} \label{damping}
\gamma(T) = \gi \left( \frac{T}{\Ti} \right) ^{4+n},
\end{equation}
where $\gi=\gamma(\Ti)$ and $n>0$.
%In some models, the exponent $n$ is related to the low relative velocity $v$ of the forward WIMPs scattering amplitude $\mathcal{M}^{\rm forward}$ off particles in the plasma,
%\begin{equation} \label{amplitude}
%|\mathcal{M}^{\rm forward}|^2 = {\rm const} \, v^n . %c\,\left(\frac{T}{M_\chi}\right)^n.
%\end{equation}
Finally, setting $\Ui=\gi/\Hi$, Eq.~\eqref{def_f} is given by
\begin{align}
\Upsilon = \frac{\gamma}{H} = \Ui  \left( \frac{T}{\Ti} \right) ^{4+n-\nu} = \Ui  \left( \frac{\ai}{a} \right) ^{\alpha(4+n-\nu)}.
\end{align}

\subsection{Kinetic temperature}

Using the definition in Eq.~\eqref{def_s} in the power-law model, we find
\begin{equation} \label{def_s}
s \equiv s(a) = \begin{cases}
\displaystyle
\frac{2\,\Ui}{\alpha(4+n-\nu)}\,\left(\frac{\ai}{a}\right)^{\alpha(4+n-\nu)},&\hbox{for $4+n \neq \nu$},\\[1ex]
\displaystyle
-2\Ui\,\ln\left(\frac{a}{\ai}\right),&\hbox{for $4+n = \nu$}.
\end{cases}
\end{equation}
%It is interesting to observe that for $4+n \neq \nu$
%\begin{align} \label{define_s}
%s = \frac{2}{\alpha(4+n-\nu)} \, \frac{\gamma}{H} .
%\end{align}
Plugging Eqs.~(\ref{H_a_powerlaw}) and~(\ref{T_a_powerlaw}) into Eq.~(\ref{sol2}),  computing the integrals,  using the identity
\begin{equation} \label{identity_gamma}
\Gamma(1+r,x) = r\,\Gamma(r,x) + x^r\,e^{-x},
\end{equation}
and defining
\begin{align} \label{define_lambda}
\lambda = \frac{2-\alpha}{\alpha\,(4+n-\nu)},
\end{align}
we find
%\begin{equation} \label{Tchi_powerlaw}
%T_\chi = \begin{cases}
%\Ti\,\left(\frac{s}{\si}\right)^{\frac{2}{\alpha(4+n-\nu)}}\,e^{s-\si}+T\,s^{\frac{2-\alpha}{\alpha(4+n-\nu)}}\,e^s\,\left[\Gamma\left(1-\frac{2-\alpha}{\alpha(4 + n - \nu)}, s\right) - \Gamma\left(1-\frac{2 -\alpha}{\alpha(4 + n - \nu)}, \si\right)\right],\quad\hbox{for $4+n \neq \nu$},\\
%\Ti\,\left(\frac{\ai}{a}\right)^{2+2\Ui} + \frac{2\Ui\,T}{2+2\Ui-\alpha}\,\left[1-\left(\frac{\ai}{a}\right)^{2+2\Ui-\alpha}\right],\quad\hbox{for $4+n = \nu$}.
%\end{cases}
%\end{equation}
\begin{equation}  \label{Tchi_powerlaw}
T_\chi = \begin{cases}
T\,s^{\lambda}\,e^s\,\big[\Gamma\left(1-\lambda, s\right) + \lambda\,\Gamma\left(-\lambda, \si\right)\big],& \hbox{for $4+n \neq \nu$},\\
\Ti\,\left(\frac{\ai}{a}\right)^{2+2\Ui} + \frac{2\Ui\,T}{2+2\Ui-\alpha}\,\left[1-\left(\frac{\ai}{a}\right)^{2+2\Ui-\alpha}\right],& \hbox{for $4+n = \nu$}.
\end{cases}
\end{equation}
To the best of our knowledge, the expressions in Eq.~\eqref{Tchi_powerlaw} have never been derived for the case of an arbitrary power-law model.

If the initial scale factor $\ai$ is taken so far back in time that the WIMPs are initially tightly coupled to the primordial plasma, then $\gi \gg \Hi$ and $\si \to +\infty$, and we obtain 
\begin{equation} \label{Tchi_powerlaw2}
T_\chi = T\,s^{\lambda}\,e^s\,\Gamma\left(1-\lambda, s\right).
\end{equation}
Eq.~(\ref{Tchi_powerlaw2}) is a generalization of the relation obtained in Ref.~\cite{bertschinger} for any cosmological power-law model and for any value of the partial wave number $n$.

\subsection{Late time behavior}

When the plasma temperature is much smaller than $T_i$, the late-time behavior of the first line of Eq.~(\ref{Tchi_powerlaw}) gives
\begin{equation} \label{Tchi_powerlaw2_noC}
T_\chi = \Ti\,s_i^\lambda\,\left(\frac{T}{\Ti}\right)^{\frac{2}{\alpha}}\,\Gamma\left(1-\lambda\right).
\end{equation}
In a cosmological model that approaches the radiation-dominated scenario where $\alpha = 1$ and $\nu = 2$, Eq.~\eqref{Tchi_powerlaw2_noC} reads
\begin{equation} \label{Tchi_stdcosm_noC}
T_\chi = \frac{T^2}{\Ti}\,\left(\frac{2\,\Ui}{2+n}\right)^{\frac{1}{2+n}}\,\Gamma\left(\frac{1+n}{2 + n}\right).
\end{equation}
We compare this result with the theoretical behavior~\cite{bringmann}
\begin{equation}\label{def_Bringmann}
T_\chi^{\rm th} = \frac{T^2}{\Tstd}\,\left(\frac{2}{2+n}\right)^{\frac{1}{2+n}}\,\Gamma\left(\frac{1+n}{2+n}\right),
\end{equation}
where $\Tstd$ is the temperature of kinetic decoupling in the radiation-dominated cosmology,
\begin{equation} \label{Tkd_std}
\Tstd = T_i\,\left(\frac{H^{\rm rad}(T_i)}{\gamma_i}\right)^{\frac{1}{2+n}},
\end{equation}
and $H^{\rm rad}(T)$ is the Hubble rate in the radiation-dominated cosmology. This latter equation can be stated in terms of the function $\Ui$ in Eq.~(\ref{Tkd_bg}) as
\begin{equation} \label{TRH_Tkd}
\Ti = \Tstd\,\Ui^{\frac{1}{2+n}}.
\end{equation}
This relation is also obtained by comparing the result in Eq.~\ref{Tchi_stdcosm_noC} with the theoretical Eq.~\eqref{def_Bringmann}. We rewrite Eq.~\eqref{TRH_Tkd} in terms of the temperature of kinetic decoupling $\Tkd$ by using the relation in Eq.~\eqref{Tkd_bg} in the form
\begin{equation}\label{Tkd_bg}
\Ui = \left(\frac{T_i}{\Tkd}\right)^{4+n-\nu},
\end{equation}
as
\begin{equation} \label{TRH_Tkd_noC}
\Tkd = \left(\frac{\Tstd^{n+2}}{\Ti^{\nu-2}}\right)^{\frac{1}{4+n-\nu}} = T_i\,\left(\frac{H^{\rm rad}(T_i)}{\gamma_i} \right)^{\frac{1}{4+n-\nu}}.
\end{equation}
Eq.~\eqref{TRH_Tkd_noC} gives the temperature of the WIMP kinetic decoupling in a generic cosmological model, which might differ from the radiation-dominated scenario at the time of decoupling. Notice that, in the particular case in which the decoupling occurs in a radiation-dominated scenario ($\nu = 2$), Eq.~\eqref{TRH_Tkd_noC} gives
\begin{equation} \label{TRH_Tkd_std}
\Tkd = \Tstd.
\end{equation}
In the following, we discuss the decoupling of WIMPs in a broken power law cosmological model, where a generic pre-BBN cosmology takes place before $\Ti$, after which standard radiation-dominated cosmology begins.

\section{Summary} \label{Summary}

In Eq.~(\ref{sol2}), we presented a general expression that gives the value of the WIMP kinetic temperature $T_\chi$ in terms of the temperature of the Universe $T$. In addition, we have presented the expression for $T_\chi$ in the case of a power-law cosmology in Sec.~\ref{Power-law models}. The expression for the temperature of kinetic decoupling in a generic cosmology is found in Eq.~\eqref{TRH_Tkd_noC}.

\newpage

%\begin{acknowledgements}
%P.G. was partially supported by NSF award PHY-1068111 and PHY-1415974.
%\end{acknowledgements}


\begin{thebibliography}{50}

\bibitem{komatsu} E.~Komatsu {\it et al.} [WMAP Collaboration], Astrophys.\ J.\ Suppl.\ {\bf 180}, 330 (2009).

\bibitem{planck}  P.~A.~R.~Ade {\it et al.}  [Planck Collaboration], arXiv:1303.5076.

\bibitem{zwicky} F.~Zwicky, Phys.\ Acta {\bf 6}, 110 (1933).

\bibitem{kolb_book} E.~W.~Kolb and M.~S.~Turner, Addison-Wesley (1990).

\bibitem{jungman1996} G.~Jungman, M.~Kamionkowski, and K.~Griest, Phys.\ Rept.\ {\bf 267}, 195 (1996) [\href{http://arxiv.org/abs/hep-ph/9506380}{hep-ph/9506.380}].

\bibitem{bertone2005} G.~Bertone, D.~Hooper, and J.~Silk, Phys.\ Rept.\ {\bf 405}, 279 (2005) [\href{http://arxiv.org/abs/hep-ph/0404175}{hep-ph/0404.175}].

\bibitem{kuhlen2012} M.~Kuhlen, M.~Vogelsberger, and R.~Angulo, Phys.\ Dark Univ.\ {\bf 1}, 50 (2012) [\href{http://arxiv.org/abs/1209.5745}{astro-ph/1209.5745}].

\bibitem{hofmann} S.~Hofmann, D.~J.~Schwarz, and H.~Stocker, Phys.\ Rev.\ D {\bf 64}, 083507 (2001)  [\href{http://arxiv.org/abs/astro-ph/0104173}{astro-ph/0104173}].

\bibitem{chen} X.~Chen, M.~Kamionkowski, and X.~Zhang, Phys.\ Rev.\ D {\bf 64}, 021302 (2001) [\href{http://arxiv.org/abs/astro-ph/0103452}{astro-ph/0103452}].

\bibitem{berezinsky} V.~Berezinsky, V.~Dokuchaev, and Y.~Eroshenko, Phys.\ Rev.\ D {\bf 68}, 103003 (2003) [\href{http://arxiv.org/abs/astro-ph/0301551}{astro-ph/0301551}].

\bibitem{green} A.~M.~Green, S.~Hofmann, and D.~J.~Schwarz, Mon.\ Not.\ Roy.\ Astron.\ Soc.\ {\bf 353}, L23 (2004); JCAP {\bf 0508}, 003 (2005) [\href{http://arxiv.org/abs/astro-ph/0503387}{astro-ph/0503387}].

\bibitem{bertschinger} E.~Bertschinger, Phys.\ Rev.\ D {\bf 74}, 063509 (2006) [\href{http://arxiv.org/abs/astro-ph/0607319}{astro-ph/0607319}].

\bibitem{bringmann} T.~Bringmann and S.~Hofmann, JCAP {\bf 0407}, 016 (2007) [\href{http://arxiv.org/abs/hep-ph/0612238}{hep-ph/0612238}].

\bibitem{kasahara} J.~Kasahara, Ph.\ D.\ dissertation, University of Utah (2009; ISBN 9781109295320; UMI microform 3368246).

\bibitem{bi} X.~J.~Bi, P.~F.~Yin, and Q.~Yuan, Phys.\ Rev.\ D {\bf 85}, 043526 (2012) [\href{http://arxiv.org/abs/1106.6027}{hep-ph/1106.6027}].

\bibitem{gondolo} P.~Gondolo, J.~Hisano, K.~Kadota, \href{http://arxiv.org/abs/1205.1914}{arXiv:1205.1914} [hep-ph].

\bibitem{visinelli} L.~Visinelli and P.~Gondolo, \href{http://journals.aps.org/prd/abstract/10.1103/PhysRevD.91.083526}{Phys.\ Rev.\ D {\bf 91} (2015) 8, 083526} [\href{http://arxiv.org/abs/1501.02233}{astro-ph/1501.02233}].

\bibitem{schmid} C.~Schmid, D.~J.~Schwarz, and P.~Widerin, Phys.\ Rev.\ D {\bf 59}, 043517 (1999) [\href{http://arxiv.org/abs/astro-ph/9807257}{astro-ph/9807257}].

\bibitem{boehm} C.~Boehm, P.~Fayet, and R.~Schaeffer, Phys.\ Lett.\ B {\bf 518}, 8 (2001) [\href{http://arxiv.org/abs/astro-ph/0012504}{astro-ph/0012504}].

\bibitem{loeb} A.~Loeb and M.~Zaldarriaga, Phys.\ Rev.\ D {\bf 71}, 103520 (2005) [\href{http://arxiv.org/abs/astro-ph/0504112}{astro-ph/0504112}].

\bibitem{profumo} S.~Profumo, K.~Sigurdson, and M.~Kamionkowski, Phys.\ Rev.\ Lett. {\bf 97}, 031301 (2006) [\href{http://arxiv.org/abs/astro-ph/0603373}{astro-ph/0603373}].

\bibitem{gondolo_gelmini} G.~B.~Gelmini and P.~Gondolo, JCAP {\bf 0810} 002, (2008) [\href{http://arxiv.org/abs/0803.2349}{astro-ph/0803.2349}].

\bibitem{aarssen} L.~G.~van~den~Aarssen, T.~Bringmann, and Y.~CGoedecke, Phys.\ Rev.\ D {\bf 85}, 123512 (2012) [\href{http://arxiv.org/abs/1202.5456}{hep-ph/1202.5456}].

\bibitem{cornell} J.~M.~Cornell and S.~Profumo, JCAP {\bf 1206}, 011 (2012) [\href{http://arxiv.org/abs/1203.1100}{hep-ph/1203.1100}].
%
%\bibitem{micromega} G.~Belanger, F.~Boudjema, A.~Pukhov, and A.~Semenov, Comput.\ Phys.\ Commun.\ {\bf 185}, 960 (2014) [\href{http://arxiv.org/abs/1305.0237}{hep-ph/1305.0237}].
%
%\bibitem{kawasaki1999} M.~Kawasaki, K.~Kohri, and N.~Sugiyama, Phys.\ Rev.\ Lett.\ {\bf 82}, 4168 (1999) [\href{http://arxiv.org/abs/astro-ph/9811437}{astro-ph/9811.437}].
%
%\bibitem{kawasaki2000} M.~Kawasaki, K.~Kohri, and N.~Sugiyama, Phys.\ Rev.\ D {\bf 62}, 023506 (2000) [\href{http://arxiv.org/abs/astro-ph/0002127}{astro-ph/0002.127}].
%
%\bibitem{hannestad2004} S.~Hannestad, Phys.\ Rev.\ D {\bf 70}, 043506 (2004) [\href{http://arxiv.org/abs/astro-ph/0403291}{astro-ph/0403.291}].
%
%\bibitem{ichicawa2005} K.~Ichikawa, M.~Kawasaki, and F.~Takahashi,  Phys.\ Rev.\ D {\bf 72}, 043522 (2005) [\href{http://arxiv.org/abs/astro-ph/0505395}{astro-ph/0505.395}].
%
%\bibitem{debernardis2008} F.~De Bernardis, L.~Pagano, and A.~Melchiorri, Astrop.\ Phys.\ {\bf 30}, 192 (2008).
%
%\bibitem{turner_LTR} M.~S.~Turner, Phys.\ Rev.\ D {\bf 28}, 1243 (1983)
%
%\bibitem{scherrer1985} R.~J.~Scherrer, M.~S.~Turner, Phys.\ Rev.\ D {\bf 31}, 681 (1985).
%
%\bibitem{dine} M.~Dine, W.~Fishler, Phys.\ Lett.\ B {\bf 120}, 137 (1983).
%
%\bibitem{steinhardt} P.~J.~Steinhardt, M.~S.~Turner, Phys.\ Lett.\ B {\bf 129}, 51 (1983).
%
%\bibitem{ford} L.~H.~Ford, Phys.\ Rev.\ D {\bf 35}, 2955 (1987).

\end{thebibliography}
\end{document}